\documentclass[twoside]{dis08}
\usepackage[latin1]{inputenc}
\usepackage[dvips]{graphicx,epsfig,color}
\usepackage{wrapfig,rotating}
\usepackage{amssymb,amsmath,array}

\begin{document}
\title{Generalized transverse momentum dependent\\
parton distributions of the nucleon}

\author{Stephan Mei{\ss}ner$^1$, Andreas Metz$^2$, and Marc Schlegel$^3$
%
%
\vspace{.3cm}\\
%
1- Institut f{\"u}r Theoretische Physik II, Ruhr-Universit{\"a}t
Bochum,\\44780 Bochum, Germany
%
\vspace{.1cm}\\
2- Department of Physics, Temple University,\\
Philadelphia, PA 19122-6082, USA
%
\vspace{.1cm}\\
3- Theory Center, Jefferson Lab, 12000 Jefferson Avenue,\\
Newport News, VA 23606, USA\\
}

\maketitle

\begin{abstract}
We present first results from our analysis of the most general quark-quark correlator of the nucleon, which can be parameterized in terms of so-called generalized transverse momentum dependent parton distributions. These results include the first complete parameterization of the nucleon GPDs and TMDs to all twists as well as new results on possible nontrivial relations between them.
\end{abstract}

\section{Introduction and definitions}
Parton distributions play an important role in the QCD description of hard scattering processes. In particular, generalized parton distributions (GPDs) and transverse momentum dependent parton distributions (TMDs) have extensively been studied during the last decade. We focus here on the most general form of quark-quark correlation functions, the so-called generalized transverse momentum dependent parton distributions (GTMDs). These GTMDs\- are connected to the so-called Wigner distributions of the hadron-parton system~\cite{Belitsky:2003nz} and contain the GPDs and TMDs in certain limits. Very recently a first complete analysis of these interesting objects has been performed, that was, however, restricted to the special case of spin-0 hadrons~\cite{Meissner:2008ay}. In this short note~\cite{URL} we will extend the analysis to the case of spin-1/2 hadrons and give a brief overview of the main applications for nucleon GTMDs.

We start by recalling some definitions, that will be needed later on. First of all, we use the common definition of the nucleon GPDs by the correlator
\begin{equation}
 F_{\lambda\lambda'}^{[\Gamma]}(x, \xi, \vec{\Delta}_T)
 = \frac{1}{2} \int\! \frac{d z^-}{2\pi} \, e^{ik  \cdot z} \,
   \big< p', \lambda' \big| \, \bar{\psi}\big(\!-\!\tfrac{1}{2}z\big) \,
   \Gamma \, \mathcal{W}_\text{GPD} \,
   \psi\big(\tfrac{1}{2}z\big) \, \big| p, \lambda \big> \, \big|_{z^+=\vec{z}_T=0} \,.
 \label{eq:corr_gpd}
\end{equation}
For $\Gamma=\gamma^+$ this correlator is parameterized in terms of two GPDs,
\begin{equation}
 F_{\lambda\lambda'}^{[\gamma^+]}(x, \xi, \vec{\Delta}_T)
 = \frac{1}{2P^+} \, \bar{u}(p', \lambda') \, \bigg[
    \gamma^+ \, H(x, \xi, t) + \frac{i\sigma^{+\mu} \Delta_\mu}{2M} \, E(x, \xi, t)
   \bigg] \, u(p, \lambda) \,,
 \label{eq:def_gpd}
\end{equation}
where $P=(p+p')/2$ denotes the average nucleon momentum and $\Delta=p'-p$ the nucleon momentum transfer. The GPDs depend on three kinematical variables: the longitudinal momentum fraction $x = k^+/P^+$ of the parton, the skewness parameter $\xi = -\Delta^+/(2P^+)$, and the momentum transfer $t = \Delta^2$. For the nucleon TMDs we also use the common definition by the correlator
\begin{equation}
 \Phi_{\lambda\lambda'}^{[\Gamma]}(x, \vec{k}_T)
 = \frac{1}{2} \int\! \frac{d z^-}{2\pi} \, \frac{d^2 \vec{z}_T}{(2\pi)^2} \, e^{ik  \cdot z} \,
   \big< P, \lambda' \big| \, \bar{\psi}\big(\!-\!\tfrac{1}{2}z\big) \,
   \Gamma \, \mathcal{W}_\text{TMD} \,
   \psi\big(\tfrac{1}{2}z\big) \, \big| P, \lambda \big> \, \big|_{z^+=0} \,.
 \label{eq:corr_tmd}
\end{equation}
Taking again $\Gamma=\gamma^+$ and introducing the nucleon spin vector $S$, which may be expressed through certain combinations of the helicities $\lambda$ and $\lambda'$, one obtains two TMDs,
\begin{equation}
 \Phi^{[\gamma^+]}(x, \vec{k}_T; S)
 = f_1(x, \vec{k}_T^2)
   - \frac{\epsilon_T^{ij} k_T^i S_T^j}{M} \, f_{1T}^\bot(x, \vec{k}_T^2) \,,
 \label{eq:def_tmd}
\end{equation}
that depend on two kinematical variables: the longitudinal momentum fraction $x = k^+/P^+$ and the transverse momentum $\vec{k}_T$ of the parton.

\section{GTMDs of the nucleon and their applications}
By examining the correlators of GPDs~(\ref{eq:corr_gpd}) and TMDs~(\ref{eq:corr_tmd}) one immediately finds that they both look very similar. Therefore, it is evident to ask, whether it is possible to study a more general correlator, that contains both in some limits. Of course this correlator is given by the canonical extension of the correlators~(\ref{eq:corr_gpd}) and~(\ref{eq:corr_tmd}) to the most general quark-quark correlator of the nucleon
\begin{equation}
 W_{\lambda\lambda'}^{[\Gamma]}(x, \xi, \vec{k}_T, \vec{\Delta}_T)
 = \frac{1}{2} \int\! \frac{d z^-}{2\pi} \, \frac{d^2\vec{z}_T}{(2\pi)^2} \, e^{ik  \cdot z} \,
   \big< p', \lambda' \big| \, \bar{\psi}\big(\!-\!\tfrac{1}{2}z\big) \,
   \Gamma \, \mathcal{W}_\text{GTMD} \,
   \psi\big(\tfrac{1}{2}z\big) \, \big| p, \lambda \big> \, \big|_{z^+=0} \,,
 \label{eq:corr_gtmd}
\end{equation}
which contains the other two correlators in the limits
\begin{align}
 F^{[\Gamma]}_{\lambda\lambda'}(x, \xi, \vec{\Delta}_T)
 &= \int\!\! d^2 \vec{k}_T \, W^{[\Gamma]}_{\lambda\lambda'}(x, \xi, \vec{k}_T, \vec{\Delta}_T) \,,
 \label{eq:lim_gpd} \\
 \Phi^{[\Gamma]}_{\lambda\lambda'}(x, \vec{k}_T)
 &= W^{[\Gamma]}_{\lambda\lambda'}(x, 0, \vec{k}_T, 0) \,.
 \label{eq:lim_tmd}
\end{align}
Note, however, that in order to have these connections, one needs to take an appropriate Wilson line $\mathcal{W}_\text{GTMD}$ in Eq.~(\ref{eq:corr_gtmd}), which reproduces the respective Wilson lines $\mathcal{W}_\text{GPD}$ in Eq.~(\ref{eq:corr_gpd}) and $\mathcal{W}_\text{TMD}$ in Eq.~(\ref{eq:corr_tmd}) in the limits above. The Wilson lines are necessary in order to ensure color gauge invariance of the correlators~(\ref{eq:corr_gpd}), (\ref{eq:corr_tmd}), and (\ref{eq:corr_gtmd}). For simplicity we restrict ourselves to the path
\begin{equation}
 -\tfrac{1}{2}z \ \ \rightarrow\ \ -\tfrac{1}{2}z + \infty \cdot n
 \ \ \rightarrow\ \ \tfrac{1}{2}z + \infty \cdot n \ \ \rightarrow\ \ \tfrac{1}{2}z
\end{equation}
or in other words to pure future-pointing ($\eta=+1$) and past-pointing ($\eta=-1$) Wilson lines. This introduces an additional dependence on $\eta = \text{sign}(n_0)$, that has been suppressed so far.

The parton distributions, which are defined by the most general quark-quark correlator of the nucleon~(\ref{eq:corr_gtmd}), are the so-called generalized transverse momentum dependent parton distribution (GTMDs). These GTMDs may be of importance for the phenomenology of hard exclusive reactions. It is already known, for instance, that gluon GTMDs enter the description of diffractive vector meson~\cite{Martin:1999wb} and Higgs production~\cite{Khoze:2000cy} even at leading twist.

The parameterization of the correlator~(\ref{eq:corr_gtmd}) can be performed by applying parity, hermiticity, and time-reversal. Analogously to Ref.~\cite{Meissner:2008ay}, where the case of a spin-0 hadron has been discussed, this yields some constraints. It follows for example that the parity constraint reduces the number of independent terms in the correlator~(\ref{eq:corr_gtmd}) significantly. Taking again $\Gamma=\gamma^+$ one obtains four complex valued GTMDs,
\begin{align}
 W^{[\gamma^+]}_{\lambda\lambda'}(x, \xi, \vec{k}_T, \vec{\Delta}_T; \eta) = \,
 &\frac{1}{2M} \, \bar{u}(p', \lambda') \, \bigg[
  F_{1,1}
  + \frac{i\sigma^{i+} k_T^i}{P^+} \, F_{1,2} \nonumber\\
 &+ \frac{i\sigma^{i+} \Delta_T^i}{P^+} \, F_{1,3}
  + \frac{i\sigma^{ij} k_T^i \Delta_T^j}{M^2} \, F_{1,4}\bigg] \, u(p, \lambda) \,,
 \label{eq:def_gtmd}
\end{align}
with $F_{1,n}=F_{1,n}(x, \xi, \vec{k}_T^2, \vec{k}_T \cdot \vec{\Delta}_T, \vec{\Delta}_T^2; \eta)$. In addition, the hermiticity constraint implies
\begin{equation}
 F_{1,n}^*(x, \xi, \vec{k}_T^2, \vec{k}_T \cdot \vec{\Delta}_T, \vec{\Delta}_T^2; \eta)
 = \pm F_{1,n}(x, -\xi, \vec{k}_T^2, -\vec{k}_T \cdot \vec{\Delta}_T, \vec{\Delta}_T^2; \eta) \,,
\end{equation}
where the plus sign holds for $n=1,3,4$ and the minus sign for $n=2$. Finally, the time-reversal constraint allows one to split each GTMD into two real valued functions,
\begin{align}
 F_{1,n}(x, \xi, \vec{k}_T^2, \vec{k}_T \cdot \vec{\Delta}_T, \vec{\Delta}_T^2; \eta) = \,
 & F_{1,n}^e(x, \xi, \vec{k}_T^2, \vec{k}_T \cdot \vec{\Delta}_T, \vec{\Delta}_T^2) \nonumber\\
 & + i \, F_{1,n}^o(x, \xi, \vec{k}_T^2, \vec{k}_T \cdot \vec{\Delta}_T, \vec{\Delta}_T^2; \eta) \,,
\end{align}
the so-called T-even and T-odd part of the GTMD, which can be distinguished by a different dependence on $\eta$. While the T-even part $F_{1,n}^e$ is independent of $\eta$ the T-odd part $F_{1,n}^o$ changes its sign under $\eta\rightarrow-\eta$.

As the correlator of GTMDs~(\ref{eq:corr_gtmd}) contains the correlators of GPDs~(\ref{eq:corr_gpd}) and TMDs~(\ref{eq:corr_tmd}) in the limits~(\ref{eq:lim_gpd}) and~(\ref{eq:lim_tmd}), GTMDs can be considered as the mother distributions of GPDs and TMDs. Therefore, one should be able to obtain the parameterization of GPDs in Eq.~(\ref{eq:def_gpd}) and of TMDs in Eq.~(\ref{eq:def_tmd}) directly from the parameterization of the GTMDs in Eq.~(\ref{eq:def_gtmd}). This is indeed possible and one finds for the GPDs
\begin{align}
 H(x, \xi, t)
 &= \int\!\! d^2 \vec{k}_T \, \bigg[ 
  F_{1,1}^e
  + 2\xi^2 \bigg( 
   \frac{\vec{k}_T \cdot \vec{\Delta}_T}{\vec{\Delta}_T^2} \, F_{1,2}^e
   + F_{1,3}^e
  \bigg)
 \bigg] \,,
 \label{eq:lim1_gpd} \\
 E(x, \xi, t)
 &= \int\!\! d^2 \vec{k}_T \, \bigg[ 
  - F_{1,1}^e
  + 2(1-\xi^2) \bigg( 
   \frac{\vec{k}_T \cdot \vec{\Delta}_T}{\vec{\Delta}_T^2} \, F_{1,2}^e
   + F_{1,3}^e
  \bigg)
 \bigg] \,,
 \label{eq:lim2_gpd}
\end{align}
with $F_{1,n}^e=F_{1,n}^e(x, \xi, \vec{k}_T^2, \vec{k}_T \cdot \vec{\Delta}_T, \vec{\Delta}_T^2)$ and for the TMDs
\begin{align}
 f_1(x, \vec{k}_T^2)
 &= F_{1,1}^e (x, 0, \vec{k}_T^2, 0, 0) \,,
 \label{eq:lim1_tmd} \\
 f_{1T}^\bot(x, \vec{k}_T^2; \eta)
 &= - F_{1,2}^o (x, 0, \vec{k}_T^2, 0, 0; \eta) \,.
 \label{eq:lim2_tmd}
\end{align}
Consequently, a complete parameterization of all GTMDs immediately yields a complete parameterization of all GPDs and TMDs. We have performed such a parameterization of the correlator~(\ref{eq:corr_gtmd}) to all twists and found a total of 64 complex valued GTMDs for the nucleon, which can be split into 64 real valued T-even and 64 real valued T-odd parts. In addition we also studied the limits of GPDs and TMDs and found that in both cases 32 real valued functions survive (see Tab.~\ref{tab}
\begin{table}[b!]
 \begin{center}{\small\begin{tabular}{|lccccc||c|c||c|c||c|c|}
  \hline
  \hspace{-.8cm} &&&&&& \multicolumn{2}{c||}{GTMDs} & \multicolumn{2}{c||}{GPDs} & \multicolumn{2}{c|}{TMDs} \\
  twist:\hspace{-.8cm} & \multicolumn{5}{c||}{$\Gamma$} & T-even & T-odd & T-even & T-odd & T-even & T-odd \\
  \hline
  2:\phantom{\Large X}\hspace{-.8cm} & $\gamma^+$\hspace{-.35cm} & /
  & \hspace{-.35cm}$\gamma^+\gamma_5$\hspace{-.35cm} & / & \hspace{-.35cm}$i\sigma^{i+}$
  & 4 / 4 / 8 & 4 / 4 / 8 & 2 / 2 / 4 & 0 / 0 / 0 & 1 / 2 / 3 & 1 / 0 / 1 \\[.05cm]
  \hline
  3:\phantom{\Large X}\hspace{-.8cm} & $1$\hspace{-.35cm} & /
  & \hspace{-.35cm}$\gamma_5$\hspace{-.35cm} & / & \hspace{-.35cm}$\gamma^i$
  & 4 / 4 / 8 & 4 / 4 / 8 & 2 / 2 / 4 & 0 / 0 / 0 & 1 / 0 / 1 & 1 / 2 / 3 \\
  \hspace{-.8cm} & $\gamma^i\gamma_5$\hspace{-.35cm} & /
  & \hspace{-.35cm}$i\sigma^{ij}$\hspace{-.35cm} & / & \hspace{-.35cm}$i\sigma^{+-}$
  & 8 / 4 / 4 & 8 / 4 / 4 & 4 / 2 / 2 & 0 / 0 / 0 & 3 / 2 / 1 & 1 / 0 / 1 \\[.05cm]
  \hline
  4:\phantom{\Large X}\hspace{-.8cm} & $\gamma^-$\hspace{-.35cm} & /
  & \hspace{-.35cm}$\gamma^-\gamma_5$\hspace{-.35cm} & / & \hspace{-.35cm}$i\sigma^{i-}$
  & 4 / 4 / 8 & 4 / 4 / 8 & 2 / 2 / 4 & 0 / 0 / 0 & 1 / 2 / 3 & 1 / 0 / 1 \\[.05cm]
  \hline
 \end{tabular}}\end{center}
 \caption{Number of independent GTMDs, GPDs, and TMDs to all twists.}\label{tab}
\end{table}
for an overview). To our knowledge, this result is new for GPDs, as so far no complete parameterization of the correlator of GPDs~(\ref{eq:corr_gpd}) has been performed to all twists. For the TMDs our result is in perfect agreement with earlier works~\cite{Goeke:2005hb}. 

Another interesting application for the GTMDs is the test of possible nontrivial relations between GPDs and TMDs. Such relations can be established in model calculations (see Ref.~\cite{Meissner:2007rx} for a review), but so far no model-independent proof has been found. The most prominent example of a possible nontrivial relation between GPDs and TMDs is probably the relation between the GPD $E$ and the Sivers function $f_{1T}^\bot$, which reads
\begin{equation}
 \int\!\! d^2\vec{b}_T \, \vec{\mathcal{I}}(x, \vec{b}_T; \eta) \,
  \frac{\epsilon_T^{ij} b_T^i S_T^{\,j}}{M} \, \mathcal{E}'(x, \vec{b}_T^{\hspace{.02cm}2})
 =- \int\!\! d^2\vec{k}_T \, \vec{k}_T \,
  \frac{\epsilon_T^{ij} k_T^i S_T^{\,j}}{M} \, f_{1T}^\bot(x, \vec{k}_T^2; \eta)
 \label{eq:nt_rel}
\end{equation}
in impact parameter space~\cite{Burkardt:2003je}. Here $\mathcal{E}$ denotes the Fourier-transformed GPD $E$ and $\vec{\mathcal{I}}$ the so-called lensing function. However, from Eqs.~(\ref{eq:lim2_gpd}) and~(\ref{eq:lim2_tmd}) it is obvious that the nontrivial relation~(\ref{eq:nt_rel}) cannot hold in general as the involved GPD and TMD have different mother distributions. The same applies to all other nontrivial relations discussed in Ref.~\cite{Meissner:2007rx}, in particular, also to the one between the GPD $\tilde{H}_T$ and the pretzelocity distribution $h_{1T}^\bot$.

\section{Summary}
In summary we have performed the first complete parameterization of the most general quark-quark correlator of the nucleon in terms of GTMDs, which can be considered as the mother distributions of GPDs and TMDs. As a by-product we also obtained the first complete parameterization of the nucleon GPDs and TMDs to all twists. This allowed us to perform a model-independent test of possible nontrivial relations between GPDs and TMDs, which can be established in model calculations. Our results indicate, that no such nontrivial relations exist as the involved GPD and TMD have different mother distributions.

\section*{Acknowledgments}
This work has partially been supported by the Verbundforschung ``Hadronen und Kerne''
of the BMBF and by the Deutsche Forschungsgemeinschaft (DFG).
\\[0.3cm]
\noindent
\textbf{Notice:} Authored by Jefferson Science Associates, LLC under U.S. DOE 
Contract No. DE-AC05-06OR23177. 
The U.S. Government retains a non-exclusive, paid-up, irrevocable, world-wide license 
to publish or reproduce this manuscript for U.S. Government purposes. 


\begin{footnotesize}




\end{footnotesize}


\end{document}